\documentclass[runningheads,a4paper]{llncs}

\usepackage[english]{babel}
\usepackage[utf8]{inputenc}
\usepackage{graphicx}
\usepackage[colorinlistoftodos]{todonotes}
\usepackage{comment}
\usepackage{mathtools}

\title{Emergence of synchrony in an Adaptive Interaction Model}

\author{Kevin Sanlaville \inst{1,2}\and Gérard Assayag\inst{1} \and Frédéric Bevilacqua\inst{1} \and Catherine Pelachaud\inst{2}}

\date{\today}

\institute{UMR STMS (Ircam, CNRS, UPMC), 1, place Stravinsky, 75004 Paris
\and Telecom-Paristech CNRS-LTCI, 37/39 rue Dareau, 75014 Paris}

\begin{document}

\maketitle

\begin{abstract}
In a Human-Computer Interaction context, we aim to elaborate an adaptive and generic interaction model in two different use cases: Embodied Conversational Agents and Creative Musical Agents for musical improvisation.
To reach this goal, we'll try to use the concepts of adaptation and synchronization to enhance the interactive abilities of our agents and guide the development of our interaction model, and will try to make synchrony emerge from non-verbal dimensions of interaction. 
\end{abstract}

\section{Introduction}

Interaction can be defined as the ensemble of reciprocal actions and responses of individuals and groups acting upon each other. It concerns verbal and nonverbal communication, implying conscious and nonconscious, enduring and casual processes. It can be considered globally as an a continually emerging process \cite{simpson1986interaction}.
This is nowadays a growing topic of interest in the fields of Computer-Human Interaction and Social Signal Processing, where the dynamics of interaction are used to perform more seamless and believable interaction between human and artificial agents. 

In this PhD project, our goal is to develop an interaction model using group dynamics.  It should be able to take into account the different modalities of interaction, especially the non-verbal communication. In particular, we aim at designing a model that can enable the emergence of synchronization in the interaction.
We expect that synchronization processes, which are a form of temporal adaptation, could make interaction dynamical in its various temporal dimensions (e.g. a conversation turn, an entire dialogue, repeated interaction, etc.). This model will then be applied to two use cases: Creative Musical Agents and Embodied Conversational Agents.

In the following section, we will try to describe the basic concepts of our interaction model. We will then present our research questions and directions.

\section{Background}

\paragraph{Synchrony and adaptation}

Modeling interaction must ideally take into account dynamic aspects and should thus be adapt to the evolution of the system. For this reason, we chose to study adaptation and one form of temporal adaptation: synchrony.
Adaptation in interaction is the way the system will evolve to match the stimuli given by the interaction context \cite{dooley1996complex}. This adaptation can be performed at various level of abstraction and timescales.
Delaherche et al. \cite{delaherche2012interpersonal} propose a definition of synchrony. They describe synchrony as a dynamical and reciprocal adaptation of the temporal structure of behaviors between interactive partners. It should be at the same time dynamical, because its main features are temporality and not actions themselves, multi-modal \cite{grammer1998courtship}, as opposite to simple imitation or the chameleon effect \cite{chartrand1999chameleon} that involve only one dimension, and happening in every interaction context, may it be cooperative or competitive. In adults, synchrony have two main roles. First, non-verbal synchrony would ease the construction of individual social connections \cite{lafrance1982posture}. Synchrony could also enhance cooperation between individuals, especially by augmenting group cohesion \cite{wiltermuth2009synchrony}.

\paragraph{Nonverbal behavior and Turn-Taking}
Nonverbal behavior can be described "as all actions distinct from speech" \cite{mehrabian1977nonverbal}, although it takes into account paralinguistic aspects of speech like prosody. Nonverbal behavior during communication can take various forms and expression supports (called modalities), such as gaze \cite{kendon1967some} or paralinguistic signals \cite{goodwin1981conversational,allwood1976linguistic}. According to Knapp et al., all human beings are natural "experts" in multimodal communication \cite{knapp2012nonverbal}, i.e. they are able to emit and receive simultaneously signals on different modalities knowingly or unknowingly. According to Argyle \cite{argyle2013bodily}, non-verbal communication fill four main functions: Express emotions, send interpersonal attitudes, present oneself's personality and accompany parole. This last notion is essential in turn-taking mechanisms. A turn in conversation can be defined as the moment between the taking of the floor to the withdrawal of the floor, that can be either consensual or forced \cite{goodwin1981conversational}. Overlapping turns can indicate a conflict between speakers but can also mean a high level of synchrony between interactants as they are able to decipher the cues of abandoning of the turn \cite{allwood1995activity}. Turn-Taking mechanisms are the way human regulates their conversational interactions and is perceivable through signals emitted by both the speakers and their interlocutors. Duncan identifies threes types of these signals in conversation \cite{duncan1972some}: turn-passing signals, signals to keep or try to take the turn from another speaker and backchannels that indicate multiple attentions, from the simple acknowledging of one's last utterance to the expression of the mental state of the emitter \cite{allwood1976linguistic}. These signals are essential to conversation and warrant its fluidity \cite{ten2005temporal}.

\section{Research Questions}

\begin{figure}
\centering
\includegraphics[width=0.2\textwidth]{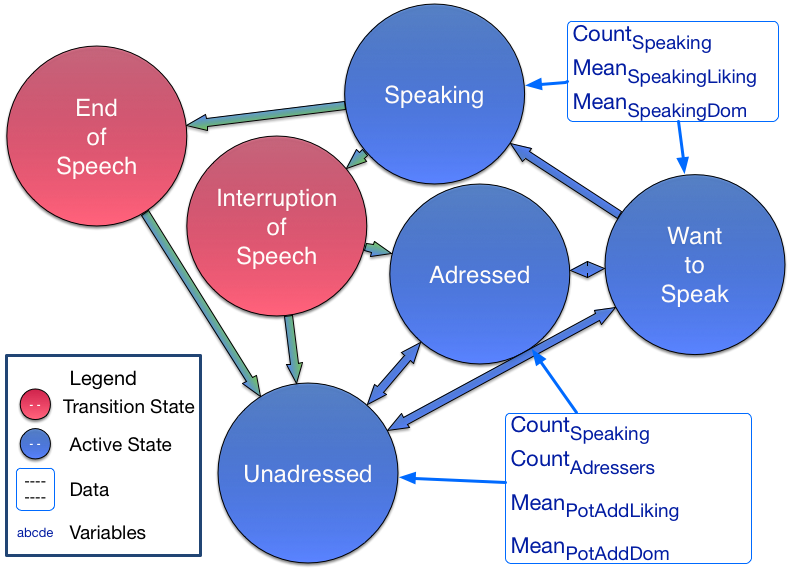}
\caption{\label{fig:TTModel}A general view of our turn-taking model.}
\end{figure}

We intend to model synchrony emergence in Turn-Taking behavior in a group of agents whether they are taking part into a cooperative or competitive interaction. Our model describes an individual and the way it perceives other agents. Each agent is able to generate a meaningful output (that from this point we will define as conversation for clarity's sake ) and non-verbal cues through body animation and para-linguistic features. The agent model is based on a Turn-Taking system by Ravenet et al. (in press), that we modified. The model in itself can be shown in Fig. \ref{fig:TTModel}.

The Turn-Taking system can be modeled as a Finite State Machine \(A = \{S,\Sigma,s_{0},\delta,F\} \) where: 
\begin{itemize}
\item \(S = \{\scriptstyle Unaddressed, Addressed, WantToSpeak, Speaking, InterruptionOfSpeech,EndOfSpeech \displaystyle \} \) the conversational states of the agent
\item \(\Sigma \) the transition matrix
\item \(s_{0} = Unaddressed\) the initial state
\item \(\delta: \Sigma \times S \to S \) the transition function
\item \(F = \{\emptyset\}\) the (empty) final states of the FSM
\end{itemize}

In this Turn-Taking system, states S describe the current mindset of the agents regarding the conversation, which could be unaddressed, addressed, wanting the turn, speaking, being interrupted and ending the speech and giving the floor to the other participants. 
Each agent does not know the exact state of conversation in which the other agents are but are able to infer it through the non-verbal cues, backchannels and speech it perceive ; for instance, in a simple dyadic use case, an agent will know it is addressed by another agent if it perceives that the other agent is speaking, that it is oriented towards the one agent and that the one agent displays cues of attention. Transitions between these states are guided by interpersonal attitude \cite{argyle2013bodily}, modeled through two dimensions: liking and dominance. Liking can be defined as “a general and enduring positive or negative feeling about some person, object or issue ” \cite{moshkina2003tameing} and dominance as “the capacity of one agent to affect the behavior of another ” \cite{prada2008social}. Interpersonal attitude is private to an agent an directed towards another agent. An example of state transition could be

\[\delta( \scriptstyle \{  Mean_{S,D}, Mean_{S,L},Count_{S} \}\displaystyle,\scriptstyle WantToSpeak \displaystyle ) =  \left\{ \begin{array}{l l l} Speaking  & \quad \text{if $Mean_{S,D} + | Mean_{S,L}| \geq 0$} \\ Speaking  & \quad \text{if $Count_{S} = 0$} \\ WantToSpeak & \quad \text{otherwise} \end{array} \right.\]

where:
\begin{itemize}
\item \(Mean_{S,D}\) is the mean of dominance values felt by the agent towards other agents speaking at this moment
\item \(Mean_{S,L}\) is the mean of liking values felt by the agent towards other agents speaking at this moment
\item \(Count_{S}\) is the number of other agents that are speaking at this moment
\end{itemize}

The dominance and the liking felt by an agent towards an agent can evolve through time. For instance, an agent interrupting another agent will feel its dominance value increase towards this agent, whereas the other agent can see its liking value decrease towards the agent that interrupted him, and since we use liking to determine the drive from an agent to speak to another, a decrease in liking will mean that the other agent will be less inclined to speak with the one who interrupted it. These values determining the Turn-Taking behavior of the system, we expect these values to converge to a defined close range, adapting to the change in the system but keeping it in a stable state and therefore making the Turn-Taking behavior synchronize. We intend to verify the existence of this synchrony in our system through the usage of automated method such as phase synchronization \cite{quiroga2002performance} or mutual information \cite{kraskov2004estimating}, and also through subjective evaluation by naive users to verify that the synchronous behavior observed in the agents are still similar to the behaviors taking place in human-human interaction. Since the agent have states that are inferred through observation, we intend to use Hidden Markov Models to describe our FSM.

\begin{figure}
\centering
\includegraphics[width=0.3\textwidth]{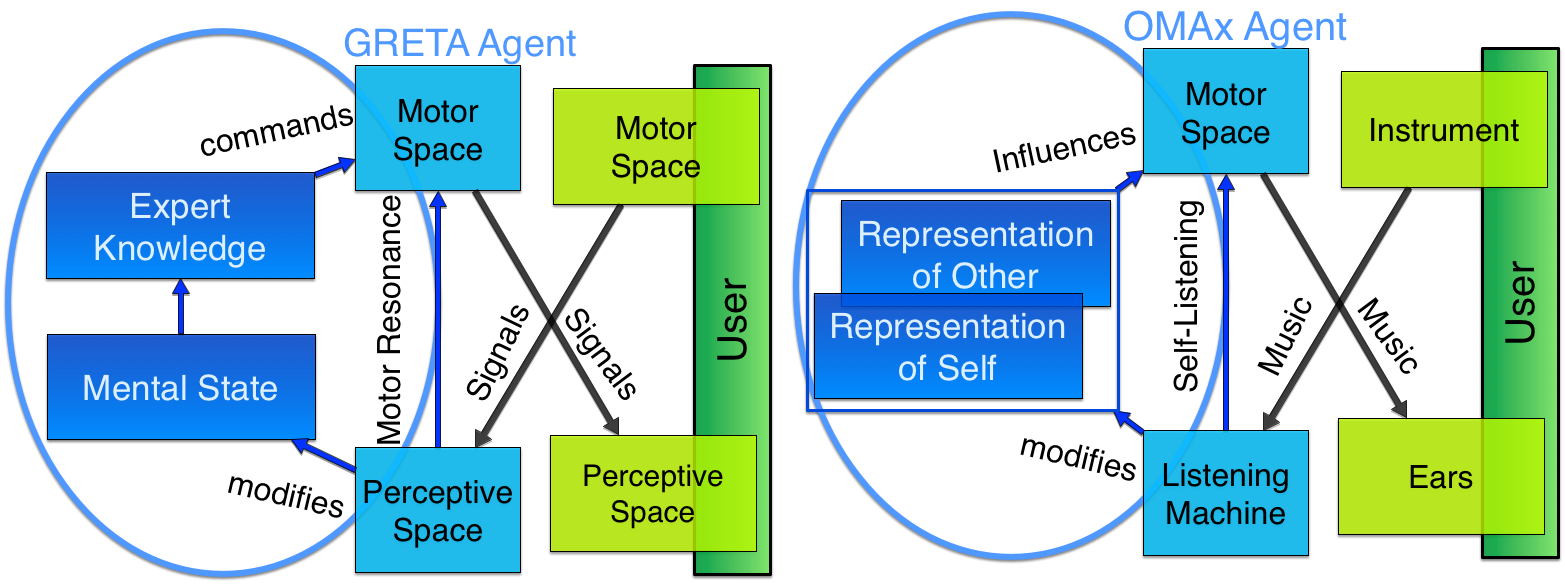}
\caption{\label{fig:GRETAOMAX}A general view of the OMAX and GRETA architectures.}
\end{figure}

\paragraph{Use cases: Existing Architectures}
The choice to pick musical improvisation as a use case was motivated by the nature of this phenomenon herself: according to Borgo et al., improvisation can be viewed as the "synchronization of our intentions and our actions, and also the upholding of a connection, a sensibility with group dynamics and evolutive experiences" \cite{borgo2004sync}.The OMax System \cite{levy2012omaxist} (See Fig \ref{fig:GRETAOMAX}, right) is an automatic improvisation mechanism that rely on the notion of stylistic reinjection \cite{assayag2006improvisation}, i.e. a system that extract characteristic elements of a musical sequence to devise a model which describe the style of the played sequence. After the listening of a musical sequence by a human instrumentalist, it can replay a similar sequence presenting stylistically close variations of what have been already played thanks to Factor Oracle \cite{allauzen1999factor}. Musical interaction between the musician and OMax is divided in two phases. In the listening phase, OMaX will perceive the musical sequence which will be decomposed note by note and stocked in the memory of the system where transitions between non-consecutive but similar states will be created thanks to the particular structure of the Factor Oracle. In the playing phase, a human operator select the sequences and sub-sequences of the memory for the system to play. If the operator select transitions between non-consecutive states, he/she introduces variety in the sequence though respecting the style of the sequences played by the human musician.


The GRETA-VIB \cite{niewiadomski2011cross} (See Fig \ref{fig:GRETAOMAX}, left) system is a virtual embodied character that uses a modular architecture independent of the agent’s embodiment . This architecture follows the SAIBA framework that specifies three modules: the intent planner, the behavior planner and the behavior realizer. The modularity is at the center of the GRETA-VIB architecture. In addition to the three modules implementing the SAIBA framework, each designer can provide the program with independent module attached to these “backbone” modules and that could specify the characteristic of the ECA, notably its behavior, independently from the way it is embodied. One of these modules implements a Turn-Taking mechanism. The Turn-Taking system is done through a Finite State Machine (FSM) that specify the current state of the agent and the different transition between states regarding whether the agent is addressed or no and the interpersonal social attitude (modeled through the dominance and liking dimensions). Each agent has by now the knowledge of the state of all the other agent, but do not know either the interpersonal attitude towards him or the other agents or the internal variables such as the number of people addressing the agent.

\section{Current and Future work}
We first established a literature basis to ground our idea of a synchronous interaction model and to apply it to our use cases. 
We are now looking to implement a first simple prototype using Hidden Markov Models or their extension called Influence Models \cite{dong2007using}. To guide the emergence of synchrony, we are now looking for metrics related to its expression and how to evaluate its occurrence, and will be very interested in every input the community can provide us on these questions. 

\section{Acknowledgments}
This work was performed within the Labex SMART (ANR-11-LABX-65) supported by French state funds managed by the ANR within the Investissements d'Avenir programme under reference ANR-11-IDEX-0004-02 and the DYCI2 Project under reference ANR-14-CE24-0002-01.

\bibliographystyle{splncs03}
\bibliography{Biblio.bib}
\end{document}